\documentclass[12pt]{article}
\usepackage{a4wide}
\usepackage{amssymb}
\usepackage{graphicx}
\begin{document}
{\renewcommand{\thefootnote}{\fnsymbol{footnote}}
\begin{center}
{\LARGE Space-time physics in background-independent theories of quantum gravity}\\
\vspace{1.5em}
Martin Bojowald\footnote{e-mail address: {\tt bojowald@gravity.psu.edu}}\\
\vspace{0.5em}
Institute for Gravitation and the Cosmos,\\
The Pennsylvania State
University,\\
104 Davey Lab, University Park, PA 16802, USA\\
\vspace{1.5em}
\end{center}
}

\setcounter{footnote}{0}

\begin{abstract}
  Background independence is often emphasized as an important property of a
  quantum theory of gravity that takes seriously the geometrical nature of
  general relativity. In a background-independent formulation, quantum gravity
  should determine not only the dynamics of space-time but also its geometry,
  which may have equally important implications for claims of potential
  physical observations. One of the leading candidates for
  background-independent quantum gravity is loop quantum gravity. By combining
  and interpreting several recent results, it is shown here how the canonical
  nature of this theory makes it possible to perform a complete space-time
  analysis in various models that have been proposed in this setting. In spite
  of the background-independent starting point, all these models turn out to
  be non-geometrical and even inconsistent to varying degrees, unless strong
  modifications of Riemannian geometry are taken into account. This outcome
  leads to several implications for potential observations as well as lessons
  for other background-independent approaches.
\end{abstract}

\section{Introduction}

A key feature of general relativity is its ability to determine both the
dynamics and the structure of space-time. A complete quantum theory of gravity
should therefore refrain from presupposing space-time structure; only then can
it be considered a proper quantization of the theory. As a conclusion,
space-time structure must be derived after quantization for a subsequent
physical analysis, and the result may be modified compared with the familiar
Riemannian structure.  Depending on the quantization procedure, it may even
happen that no consistent space-time structure exists for its solutions. A
detailed analysis is then required to see whether the theory can be considered
a valid candidate for quantum gravity, even if it is formally consistent,
judged by non-geometrical standards such as conditions commonly imposed on
quantizations of gauge theories.  These questions are highly non-trivial in
any approach. A detailed analysis is now available in models of loop quantum
gravity, but it remains preliminary owing to the tentative nature of physical
models of space-time in this theory.

Loop quantum gravity is often advertized as a background-independent approach
to quantum gravity. This characterization suggests that the theory might
indeed be free of pre-supposed space-time structures. In practice, however,
the rather involved nature of methods suitable for derivations of space-time
structures, combined with the canonical treatment used in the more successful
realizations of loop quantum gravity, has for some time obscured the role and
nature of space-time in this theory. In fact, several long-standing doubts
exist as to the possibility of covariance in models of loop quantum
gravity. For instance, the ``bounce'' idea, used in a majority of cosmological
and black-hole models in this setting, is largely based on calculations
available for the dynamics in homogeneous cosmological models, introducing
formal properties of discreteness or boundedness seen in the kinematics of the
full theory of loop quantum gravity. Since it remains unknown whether there is
space-time dynamics consistent with the kinematics of the full theory,
there is no guarantee that kinematical ingredients exported to homogeneous
models of quantum cosmology can give rise to a meaningful structure of
space-time and some sense of general covariance.

More specifically, kinematical features that apply spatial discreteness work
like a cut-off which, if it is a fixed scale, is hard to reconcile with the
transformations required for covariance. If one accepts the possibility that
quantum gravity may well lead to non-classical space-time structures that
require a modified and perhaps weakened version of general covariance,
consistency requires a detailed demonstration of how one can avoid various
low-energy problems that may then trickle down from the Planck regime, as
pointed out in \cite{LorentzFineTuning,SmallLorentzViol}. Moreover, in such a
situation it is important to determine how a modified space-time structure can
be described in meaningful terms, for instance by addressing the question of whether
such a theory can still be considered geometrical and whether there is an
extended range of parameters (such as $\hbar$) in which effective line
elements may still be available.

A consideration of space-time structure in bounce models also raises the
question of how exactly singularity theorems are evaded. In models of loop
quantum cosmology, bounce solutions are obtained without modifying matter
Hamiltonians. The standard energy conditions therefore remain satisfied,
obscuring the possibility of avoided singularities often claimed in this
setting. Since singularity theorems make statements about boundaries of
space-time and use general properties of Riemannian geometry such as the Ricci
curvature and the geodesic deviation equation, they depend on and require a
consistent form of space-time structure. Unfortunately, however, bounce models
of loop quantum gravity are often accompanied by poorly justified and
contradictory statements about space-time. For instance, standard line
elements are commonly used to express modified gravitational dynamics in
tractable form, implicitly presupposing that space-time remains
Riemannian. But then, singularity theorems should be applicable to the
resulting modified solutions since the behavior of matter energy is assumed
unchanged, making it impossible to evade singularities by a bounce. (The
behavior of singularities may depend on a possibly modified relationship
between stress-energy and Ricci curvature even if one maintains
positive-energy conditions. However, simple bounce models based on modified
Friedmann equations do not provide such a relationship because their
space-time structure remains unclear.)  The fact that this contradiction has
gone unnoticed for several years in this field serves to highlight the
challenging nature of questions about space-time in loop quantum gravity.

Independently of bounce claims, results about space-time structure in models
of loop quantum gravity have been accumulating in recent years.  This review
presents a summary, highlighting similarities between different ways in which
covariance can be and often is violated. By now, all the high-profile claims
made in the last decade in the context of loop quantum gravity, including
\cite{AANLetter,LoopSchwarz,LoopCollapse,Transfig}, have been shown to rest on
inconsistent assumptions about space-time structure and covariance. It is
therefore of interest to combine and compare the various ways in which
covariance can be violated in order to arrive at a general perspective. (Some
of these models have already been presented in an overview form in
\cite{BlackHoleModels}. The focus of this previous review was on implications
for models of black holes, while the present one emphasizes the role of these
results for general aspects of background independence and the viability of
quantum gravity. Moreover, it presents further comparisons between the
different results.)

A discrete fundamental theory is not expected to respect all the properties
that we are used to from classical space-time. Some violation of classical
covariance may therefore be allowed. Nevertheless, because covariance not only
describes a property of classical space-time but also implies that all
consistency conditions are met for gravity as a gauge theory, the requirement
of general covariance cannot just be abandoned without suitable
replacements. One task to be completed for a consistent theory of quantum
gravity is to find suitable middle ground between completely broken covariance
and the strictly classical notion of general covariance. Considerations of
covariance therefore remain important even if one believes that quantum
gravity may completely change the structure of space-time in its fundamental
formulation.

The examples of violations of covariance discussed here do not directly apply
to fundamental quantum gravity but rather to models used for phenomenological
studies of cosmology or black holes. In this context, the question of
covariance is even more pressing because a general (but often implicit)
strategy in this context is to use well-understood Riemannian geometry to
analyze potential modifications in the dynamical equations of quantum
gravity. Since these modifications may easily affect space-time structure as
well, any implicit assumptions about space-time must be uncovered and analyzed
before an analysis can be considered meaningful. In this phenomenological
context, the question of space-time structure is not as challenging as it is
at the fundamental level, but it is still relevant. The task is to show that a
certain geometrical structure applies to solutions of an effective description
of quantum gravity not only in the strict classical limit where $\hbar=0$ but
also within some finite range of the expansion parameter, given for instance
by $\rho/\rho_{\rm P}$ in a cosmological model with energy density $\rho$
relative to the Planck density.

The studies \cite{AANLetter,LoopSchwarz,LoopCollapse,Transfig} of interest
here implicitly assume that space-time structure remains unmodified even in
the presence of modified dynamics, and sometimes even all the way to the
Planck scale \cite{AANLetter,Transfig}. This strong assumption is implemented
by inserting solutions of modified equations in a standard line element,
without checking whether the modified solutions obey gauge transformations
compatible with coordinate transformations such that an invariant line element
results. Such a line element is crucial in these studies because it made it
possible to formulate new claims of potential physical effects that made these
studies interesting and publishable in high-profile journals. The same
ingredient makes these studies vulnerable to violations of covariance, as
reviewed in detail in the following sections.

The concluding section of this review points out general properties of
covariance in models of loop quantum gravity that may be useful for other
approaches. It is generally expected that quantum gravity leads to new
geometrical features at large curvature that can no longer be described by a
classical form of space-time with its common sense of covariance. Loop quantum
gravity is only one approach in which a specific example of discreteness or
other non-classical geometrical effects is being explored. The general
question to be addressed is then whether quantum gravity at large curvature
remains a geometrical theory in the sense that its solutions can still be
described as space-times with a certain generalized meaning compared with our
classical notion.

\section{Models of loop quantum gravity}

In order to set up our analysis, we should first introduce the general form of
modifications implemented in models of loop quantum gravity. (See \cite{ROPP}
for more details.)  It is sufficient to illustrate these modifications by
recalling the basics of loop quantum cosmology for spatially flat, isotropic
models.

\subsection{Holonomy modifications and space-time structure}

The classical dynamics of the scale factor $a$ can be expressed by a
canonical pair $(q,p)$ where $q=\dot{a}$ (a proper-time derivative) and
$|p|=a^2$, subject to the Friedmann constraint
\begin{equation}
 -\frac{{q}^2}{|{p}|}+ \frac{8\pi G}{3}
 \rho=0
\end{equation}
with the energy density $\rho$. Kinematical aspects of loop quantization
suggest the replacement, or ``holonomy modification,''
\begin{equation}
 \frac{q^2}{|{p}|}\mapsto \frac{\sin(\ell
 q/\sqrt{|{p}|})^2}{\ell^2}
\end{equation}
where $\ell$ is a suitable, possibly running length scale, such as the Planck length $\ell_{\rm
  P}$ in simple cases. 

Taken in isolation, holonomy modifications imply non-singular behavior in
isotropic models with a modified Friedmann constraint
\begin{equation} \label{HolFriedmann}
 \frac{\sin(\ell q/\sqrt{|{p}|})^2}{\ell^2}= \frac{8\pi
   G}{3} \rho
\end{equation}
because the energy density of any solution to this equation must be bounded (assuming that
$\ell$ is constant, as commonly done in this context). However, this equation
includes only one type of expected quantum corrections. In addition, a
complete effective description of some underlying dynamics of quantum gravity
(of any kind) should also include the remnants of higher-curvature terms in an
isotropic model. Higher-curvature terms, just like holonomy modifications,
require a given length scale, which we may assume to equal $\ell$ if holonomy
modifications and higher-curvature terms are derived from a single quantum
theory of gravity. It is easy to see that higher-curvature terms are not
described by (\ref{HolFriedmann}) because they generically imply higher time
derivatives and therefore extend the phase space by additional momenta.

The equation (\ref{HolFriedmann}) is therefore incomplete from the viewpoint
of effective theory. Nevertheless, it may be useful because it determines at
least one type of quantum corrections. However, knowing that there are
additional terms not included in (\ref{HolFriedmann}) that also depend on
$\ell$, we cannot trust the full function $\sin^2(\ell q/\sqrt{|p|})/\ell^2$
but should rather expand
\begin{equation}
 \frac{\sin(\ell
 q/\sqrt{|{p}|})^2}{\ell^2}\sim \frac{q^2}{|{p}|}\left(1-
 \frac{1}{3}\ell^2\frac{q^2}{|{p}|}+\cdots\right)
\end{equation}
and include only leading-order terms.  If $\ell\sim \ell_{\rm P}$, these
leading corrections are of the order $\ell_{\rm P}^2q^2/|{p}|\sim
\rho/\rho_{\rm P}$, which is the same as the order expected for
higher-curvature terms. Even the leading corrections in (\ref{HolFriedmann})
should therefore be taken with a grain of salt. Interpreting the full series
expansion or its sum to the sine function as an indication of bounded
densities is unjustified in the absence of information about
higher-curvature terms.

Higher-curvature terms are also of interest from the point of view of
space-time structure. We have already used the fact that they generically
include higher time derivatives, but the specific appearance of such terms is
not arbitrary and instead guided by requirements of general covariance. In
loop quantum cosmology, the form of quantum corrections that may appear in
addition to holonomy modifications can therefore be determined only if there
is good control on space-time structure in this setting.

Isotropic and homogeneous models are not sufficient for an analysis of
space-time structure and covariance because these questions rely on how
spatial and temporal dependencies are related in differential equations and
their solutions. At least one spatial direction of inhomogeneity should then
be included in suitable models, in addition to the non-trivial time dependence
already described by models such as (\ref{HolFriedmann}). While such
(midisuperspace) models have been considered in loop quantum gravity for some
time, their application to the question of covariance is rather new and has
led to several surprising results.

\subsection{Three examples and one theorem}

We will review three examples of proposed methods to describe inhomogeneity in
models of loop quantum gravity and the reasons why they turn out to violate
covariance in ways that render them inconsistent. The first example, the
dressed-metric approach for cosmological inhomogeneity \cite{AAN}, has been
used several times as a crucial ingredient in cosmological model building,
leading to claims of observational testability that, given the underlying
problems with space-time structure, turn out to be unfounded. (Similar
arguments regarding violations of covariance apply to the hybrid approach to
inhomogeneity in loop quantum cosmology \cite{Hybrid}.) The remaining two
examples, given by partial Abelianizations of constraints in spherically
symmetric models \cite{LoopSchwarz} as well as a misleadingly named
``covariant polymerization'' \cite{CovPol} in related studies apply to
proposed scenarios for quantum black holes. (The proposal of \cite{CovPol} was
intended to justify modified equations used for a study of critical collapse
in \cite{LoopCollapse}.)

In addition, we will describe a detailed no-go theorem based on a
minisuperspace description of the static Schwarzschild exterior by a
homogeneous timelike slicing, as originally proposed for a different purpose
in \cite{Transfig}.

\section{Dressed-metric approach}

In classical gravity, as is well known, it is possible to describe
cosmological inhomogeneity in the early universe as a coupled system of two
independent sets of degrees of freedom, given by inhomogeneous perturbations
evolving on a homogeneous background with a choice of a time coordinate (such
as proper time or conformal time). In a discussion of possibly modified
dynamics and space-time structure, it is important to remember that these two
ingredients, background and perturbations, have rather different properties
related to covariance.

\subsection{Background and perturbations}

The dynamics of any homogeneous background can be modified without violating
covariance because there is a single constraint, (\ref{HolFriedmann}), which
is always consistent with itself in any modified form: Because $\{C,C\}=0$ for
any Poisson bracket, Hamilton's equations generated by a constraint $C$ are
guaranteed to preserve the constraint equation $C=0$ imposed on initial
values.

Applied to the Friedmann constraint $C$, we generate equations of motion
\begin{equation}
 \frac{{\rm d}f}{{\rm d}t}=\{f,NC\}
\end{equation}
for any phase-space function $f$, with respect to a time coordinate $t$
indirectly determined by the lapse function $N>0$.  The generic time
derivative, applied to solutions of the constraint $C=0$, can be rewritten as
\begin{equation}
 \frac{1}{N}\frac{{\rm d}f}{{\rm d}t}=\{f,C\}=\frac{{\rm d}f}{{\rm d}\tau}
\end{equation}
introducing proper time $\tau$ in the last step by the usual definition ${\rm
  d}\tau=N{\rm d}t$.

All allowed choices of time
coordinates (monotonically related to $\tau$) can therefore be described by a
single line element,
\begin{equation}
 {\rm d}s^2 = -{\rm d}\tau^2+ \tilde{a}(\tau)^2 {\rm d}\sigma^2
\end{equation}
where $\tilde{a}(\tau)$ denotes the scale factor subject to potentially
modified dynamics, and ${\rm d}\sigma^2$ is a standard isotropic spatial line
element.  Because the definition of $\tau$ implies that the line element is
correctly transformed to
\begin{equation} \label{dsN}
 {\rm d}s^2 = -N^2{\rm d}t^2+ \tilde{a}(t)^2 {\rm d}\sigma^2
\end{equation}
for any other time coordinate $t$, there is a suitable way to describe any
modified homogeneous dynamics, subject to a single constraint, by a space-time
geometry that is invariant with respect to the full coordinate changes allowed
by the symmetry, given by reparameterizations of time.

Coordinate changes are more involved in the case of spatial inhomogeneity because
several independent coordinates may be related by transformations. In the
canomical language of constraints, the presence of a multitude of independent
ones, one Hamiltonian constraint per spatial point as well as diffeomorphism
constraints, implies that a modification of one or more constraints no longer
implies consistency of their Hamiltonian flows with respect to the other
constraints. Since the relevant constraints implement space-time
transformations, a dedicated space-time analysis then becomes important.

For small, perturbative inhomogeneity, there is a standard way to describe
curvature perturbations in terms of combinations of metric and matter fields
that are invariant with respect to small coordinate changes
\cite{Bardeen}. However, compared with reparameterizations of time relevant
for the background, it is much harder to derive a suitable invariant line
element extending (\ref{dsN}) in a way that is consistent with Hamilton's
equations generated by modified constraints for perturbative inhomogeneity. In
fact, the standard derivations of curvature perturbations
\cite{Bardeen,CosmoPert} as well as the canonical version given in
\cite{HamGaugePert} assume that space-time is of its classical form, for instance
by directly working with coordinate substitutions in a line element. A
modified or quantum treatment then cannot take it for granted that the form of these curvature
perturbations remains unchanged, because space-time structure itself may be modified
in quantum gravity.

The dressed-metric approach proceeds by quantizing standard curvature
perturbations on a modified background, leading to wave equations for
perturbations on a modified background line element ${\rm
  d}s^2=\tilde{g}_{\alpha\beta}{\rm d}x^{\alpha}{\rm d}x^{\beta}$ of the form
(\ref{dsN}). The approach therefore implicitly assumes that space-time
structure remains classical even while the dynamics at least of the background
are modified. Upon closer inspection, this assumption turns out to be
unjustified.

\subsection{The metric's new clothes}

As pointed out already in \cite{Stewart}, Bardeen variables or curvature
perturbations are ``gauge invariant'' under small coordinate changes, but not
necessarily under all coordinate changes relevant for a given cosmological
situation. In particular, in cosmological models of perturbative inhomogeneity
we also need invariance under potentially large background transformations of
time, such as transforming from proper time to conformal time.

Small coordinate changes of perturbations and large reparameterizations of
background time are not independent of each other. Algebraically, they form a
semidirect product rather than a direct one, as shown in \cite{NonCovDressed}.
The non-trivial interplay between these transformations can be deduced
from vector-field commutators such as
\begin{equation} \label{fxi}
 \left[f(t)\frac{\partial}{\partial t}, \xi^{\alpha}\frac{\partial}{\partial
     x^{\alpha}}\right]= f\dot{\xi}^{\alpha} \frac{\partial}{\partial
   x^{\alpha}}- \dot{f}\xi^0 \frac{\partial}{\partial t}
\end{equation}
which in general are not zero (in contrast to what a direct product would
imply) but rather form a small inhomogeneous transformation.
This interplay is a general property of perturbations in Riemannian geometry,
as encoded in line elements suitable for perturbative inhomogeneity.  

The applicability of standard line elements and requires the precise algebra
of coordinate transformations to be modeled by gauge transformations in a
canonical formulation of any gravity theory. However, while the dressed-metric
approach assumes the availability of standard line elements with the usual
coordinate dependence (but possibly modified metric coefficients), it violates
the algebraic condition by its independent treatment of background and
perturbations: Quantizing the background seperately from perturbations
evolving on it implicitly assumes a direct product of coordinate changes.
Writing a line element
${\rm d}s^2=\tilde{g}_{\alpha\beta}{\rm d}x^{\alpha}{\rm d}x^{\beta}$ based on
modified metric components $\tilde{g}_{\alpha\beta}$ in a dressed-metric model
is therefore meaningless.

\subsection{Effective line element}

Because a line element
${\rm d}s^2=g_{\alpha\beta}{\rm d}x^{\alpha}{\rm d}x^{\beta}$ is defined as
the square of an infinitesimal distance, it can be meaningful as a description
of geometry only if it is independent of coordinate choices that affect
${\rm d}x^{\alpha}$ as well as $g_{\alpha\beta}$. For ${\rm d}s^2$ to be
invariant, the metric coefficients $g_{\alpha\beta}$ must be subject to the
standard tensor-transformation law
\begin{equation}
 g_{\alpha'\beta'}= \frac{\partial x^{\alpha}}{\partial x^{\alpha'}}
 \frac{\partial x^{\beta}}{\partial x^{\beta'}} g_{\alpha\beta}
\end{equation}
if coordinates $x^{\alpha}$ are transformed to $x^{\alpha'}$.

Canonical quantization in its usual form, as applied in models of loop
quantum gravity, does not modify space-time coordinates $x^{\alpha}$ and their
transformations, but it may alter equations of motion (with respect to these
coordinates) for the spatial metric $q_{ij}$ in the generic canonical line element
\begin{equation}
 {\rm d}s^2 = -N^2{\rm d}t^2+ q_{ij}({\rm d}x^i+M^i{\rm d}t)({\rm
   d}x^j+M^j{\rm d}t)\,.
\end{equation}
Modifications of the remaning components, the lapse function $N$ and shift
vector $M^i$, are also determined by canonical equations, although more
indirectly because $N$ and $M^i$ do not have unconstrained momenta.  In the
presence of modifications, altered equations for $q_{ij}$, $N$ and $M^i$ must
remain consistent with coordinate transformations if an effective line element
${\rm d}s^2$ is to be meaningful.

A crucial ingredient in a canonical analysis of covariance is therefore given
by the transformations of $N$ and $M^i$, in addition to the more obvious
transformations of $q_{ij}$. The full set of canonical transformations makes
use of specific properties of the constraints of the theory. At this point,
the analysis of geometrical properties relevant for effective line elements
benefits from a discussion of hypersurface deformations in space-time, which
are generated by the constraints. While properties of hypersurface
deformations constitute some of the classic results in canonical general
relativity
\cite{Regained,LagrangianRegained,KucharHypI,KucharHypII,KucharHypIII}, they
do not appear to be widely know. What follows is a construction of
hypersurface deformations based on elementary properties of special
relativity.

\begin{figure}
\begin{center}
\includegraphics[width=12cm]{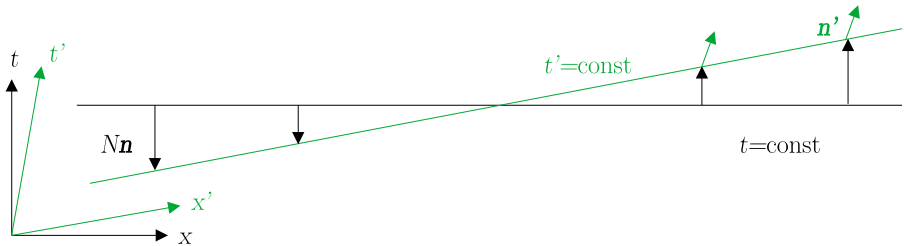}
\caption{A Lorentz transformation in Minkowski space-time, shown in the
  traditional way by means of axes as well as in terms of linear normal deformations
  of a spatial slice. A slice $t={\rm const}$ in the original coordinate
  system is transformed to a new spatial slice $t'={\rm const}$ by a linear
  deformation with position-dependent displacement $N(x)=N_0+vx$ along the
  unit normal vector field  $\bf{n}$. 
  \label{f:LorentzMink}}
\end{center}
\end{figure}

\subsubsection{Hypersurface deformations}

In special relativity, an observer moving at speed $v$ assigns
new coordinates to events in space-time according to a Lorentz transformation
\begin{equation}
 x'=\frac{x-vt}{\sqrt{1-v^2}}\quad,\quad
 t'=\frac{t-vx}{\sqrt{1-v^2}}\,.
\end{equation}
Interpreting this transformation as a linear deformation of axes in a
space-time diagram, as shown in Fig.~\ref{f:LorentzMink}, the set of all
Poincar\'e transformations can be represented geometrically by linear
hypersurface deformations with respect to lapse functions
$N(\bf{x})=\Delta t+\bf{v}\cdot \bf{x}$ (deformations in the normal direction
of a spatial slice) and shift vector fields
$\bf{M}(\bf{x})= \Delta\bf{x}+{\bf R}\bf{x}$ (tangential deformations within a
spatial slice). The parameters in these expressions for linear lapse functions
and shift vector fields determine a time translation $\Delta t$, a boost velocity
$\bf{v}$, a spatial shift $\Delta\bf{x}$ and a spatial rotation matrix
${\bf R}$.

\begin{figure}
\begin{center}
\includegraphics[width=12cm]{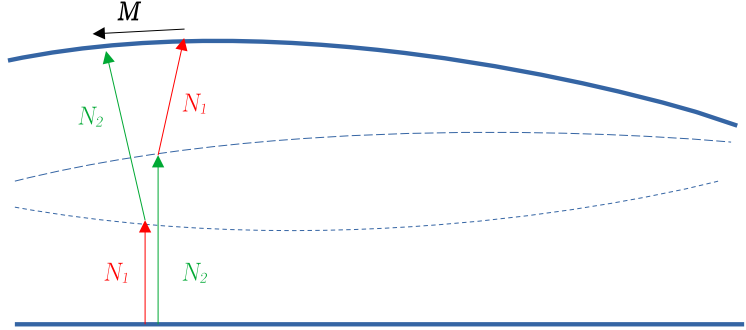}
\caption{Two non-linear normal deformations, one with a lapse function $N_1$
  and one with lapse function $N_2$, applied in two different orderings, show
  the commutator (\ref{TT}) given by a spatial displacement $\bf{M}$.
  \label{f:SurfaceDefMink}}
\end{center}
\end{figure}

We extend these considerations to general relativity by replacing the
restricted set of translations, rotations and Lorentz boosts with arbitrary
non-linear coordinate changes. Correspondingly, hypersurfaces are subject to
non-linear deformations \cite{Regained}.  Infinitesimal hypersurface deformations in Riemannian
space-time, split into ``temporal'' deformations $T(N)$ in a normal direction and
``spatial'' deformations $S(\bf{M})$ in tangential directions,
can be shown to obey the commutators
\begin{eqnarray}
 [S(\bf{M}_1),S(\bf{M}_2)]&=& S((\bf{M}_1\cdot\nabla)\bf{M}_2- 
(\bf{M}_2\cdot\nabla)\bf{M}_1) \label{SS}\\
{} [T(N),S(\bf{M})] &=& -T(\bf{M}\cdot\bf{\nabla}N)\\
{} [T(N_1),T(N_2)] &=& S(N_1\nabla N_2-N_2\nabla N_1) \label{TT}
\end{eqnarray}
when they are applied in two alternative orderings. A visualization is shown
in Fig.~\ref{f:SurfaceDefMink}.  The brackets (\ref{SS})--(\ref{TT}) represent
general covariance in canonical form. While specific expressions for $S$ and
$T$ can vary depending on the gravitational theory, such as different
higher-curvature actions \cite{HigherCurvHam}, the brackets remain the same as long as the
underlying geometry of space-time is Riemannian. Conversely, deviations of the
brackets from their Riemannian form can be used to detect non-classical
space-time structures in modified canonical gravity. The algebraic nature of
the brackets makes it possible to analyze gravitational theories without
presupposing specific geometrical formulations of space-time, constituting a
major strength of the canonical approach.

\begin{figure}
\begin{center}
\includegraphics[width=12cm]{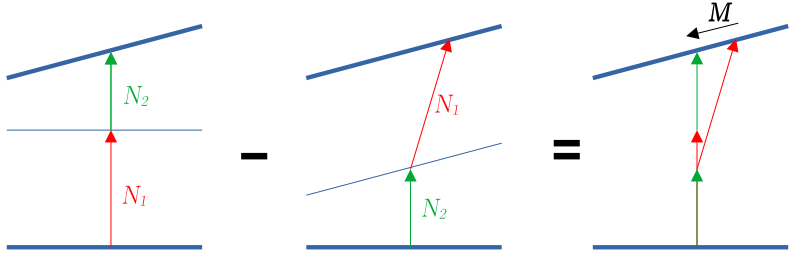}
\caption{Semidirect product of time reparameterizations and inhomogeneous
  transformations as in (\ref{fxi}), represented in the picture of
  hypersurface deformations: The commutator of two such normal deformations
  produces a non-zero spatial shift ${\bf M}$.
  \label{f:SemiDir}}
\end{center}
\end{figure}

Figure~\ref{f:SemiDir} represents the commutator of an infinitesimal time
translation and an infinitesimal normal deformation. This picture can be
interpreted as a version of the vector-field commutator (\ref{fxi}) of a
background transformation and a small perturbative transformation. The
non-zero result of (\ref{fxi}) corresponds to the presence of a spatial shift
on the right-hand side of Fig.~\ref{f:SemiDir}. Even though there is no
immediate time dependence of the canonical data on which a background vector
field as in (\ref{fxi}) would act, the semidirect product of background and
perturbative transformations is clear.  In canonical language, the failure of
the dressed-metric approach to realize the correct semidirect product means
that there is no common $T(N)$ for background and perturbations in this
setting. The non-existence of consistent temporal deformations signals the
break-down of space-time and covariance.

\subsubsection{Structure functions}

The brackets of hypersurface deformations have structure functions because the
gradient in (\ref{TT}) requires the use of the spatial metric, and therefore
depend on the geometry described by these brackets. A canonical realization of
these brackets is given by the Hamiltonian and diffeomorphism constraints,
$H[N]$ and $D[M^i]$, of a given gravity theory. Written in the form
\begin{eqnarray}
  \{D[M_1^i],D[M_2^j]\} &=& D[[M_1,M_2]^i]\\
 \{H[N],D[M^i]\} &=& -H[M_1^i\nabla_iN]\\
 \{H[N_1],H[N_2]\} &=& D[q^{ij}(N_1\nabla_jN_2-N_2\nabla_jN_1)]\,, \label{HHq}
\end{eqnarray}
they make the appearance of structure functions explicit, depending on the
inverse spatial metric $q^{ij}$. Formally, we may write the constraint
brackets as $\{C_A,C_B\}= F_{AB}^D C_D$ with indices $A$, $B$ and $D$ that
combine spatial positions with the type of the constraint (Hamiltonian or a
component of the diffeomorphism constraint). The coefficients $F_{AB}^D$ are
not constants but phase-space functions.

The presence of structure functions causes long-standing problems in the
quantization of canonical gravity \cite{Komar,NonHerm}: Upon quantization,
$q^{ij}$ as well as $D$ and $H$ are turned into operators. Maintaining closed
brackets therefore requires specific ordering, regularization, or other
choices.  Even if the brackets can remain closed under certain conditions,
quantized structure functions may be quantum corrected. A question relevant
for covariance is then whether a meaningful interpretation of the generators
as hypersurface deformations in space-time still exists.

As shown in \cite{EffLine}, a meaningful space-time interpretation does exist
at least in some cases of modified structure functions. To see this, it is
necessary to construct a space-time line element that is consistent with the
modified gauge transformations generated by (\ref{HHq}) with quantum corrected
structure functions. If these functions are modified, so are the version of
hypersurface deformations they represent, and therefore the objects $q_{ij}$,
$N$ and $M^i$ in which the brackets are formulated do not directly define the
components of a meaningful line element because this notion is based on
classical space-time with standard hypersurface deformations. However, in some
cases, suitable redefinitions of the canonical fields are available that can
serve this purpose.

A derivation of proper effective line elements is based on the general
property of Hamiltonian and diffeomorphism constraints as generators of
evolution equations, giving the time derivative
\begin{equation} \label{fdot}
 \dot{f}={\cal L}_tf = \{f,H[N]+D[M^i]\}
\end{equation}
of any phase-space function $f$ with respect to the time-evolution vector
field $t^{\alpha}=N n^{\alpha}+M^{\alpha}$. (The space-time vector field
$M^{\alpha}$ is the push-forward of the spatial vector field $M^i$ by the
embedding map of a spatial slice in space-time.) In addition, the constraints
generate gauge transformations
\begin{equation} \label{fdelta}
\delta_{\epsilon}f = \{f,H[\epsilon]+D[\epsilon^i]\}
\end{equation}
that would correspond to coordinate changes generated by the vector field
$\xi^{\alpha}=\epsilon n^{\alpha}+\epsilon^{\alpha}$ if structure functions were
unmodified. 

In all cases --- modified and unmodified structure functions --- evolution
equations and gauge transformations must be consistent with each other: A
gauge-transformed $f$ must evolve according to the general equation
(\ref{fdot}) with the same generators $H$ and $D$ as the original $f$, but
possibly with a new time-evolution vector field. Since the direction of the
time-evolution vector field within a given theory is determined by lapse and
shift, this consistency condition can be used to derive gauge transformations
for $N$ and $M^i$. Together with the gauge transformations of $q_{ij}$,
directly determined by (\ref{fdelta}) because $q_{ij}$ are phase-space
functions, all components of a candidate space-time line element can therefore
be transformed unambiguously.

For generic structure functions $F_{AB}^D$, evolution and gauge
transformations are consistent with each other provided the multipliers
$(N^A)=(N,M^i)$ gauge transform according to \cite{CUP}
\begin{equation} \label{deltaN}
 \delta_{\epsilon}N^A= \dot{\epsilon}^A+ N^B\epsilon^C F_{BC}^A\,.
\end{equation}
Unlike in the case of
$\delta_{\epsilon}q_{ij}=\{q_{ij},H[\epsilon]+D[\epsilon^i]\}$, the structure
functions appear explicitly in (\ref{deltaN}).  Structure functions, and their
possible modifications, are therefore directly relevant for space-time
structure and the existence of meaningful effective line elements
\begin{equation}
 {\rm d}s^2 = -\tilde{N}^2{\rm d}t^2+ \tilde{q}_{ij}({\rm
   d}x^i+\tilde{M}^i{\rm d}t)({\rm 
   d}x^j+\tilde{M}^j{\rm d}t)
\end{equation}
which may require field redefinitions of $\tilde{N}$, $\tilde{M}^i$ as well as
$\tilde{q}_{ij}$ if the structure functions $F^A_{BC}$ are modified.

\subsection{Lessons from hypersurface deformations}

In canonical models of modified gravity, control on space-time structure
requires full expressions for the Hamiltonian constraint $H[N]$ and the
diffeomorphism constraint $D[M^i]$ with closed brackets. This condition is
violated in the dressed-metric approach (as well as in hybrid loop quantum
cosmology) because the independent treatment of remnant coordinate freedom in
background and perturbations, the former through deparameterization and the
latter by using curvature perturbations, precludes the construction of joint
constraints for both sets of degrees of freedom. The common assumption that
space-time in this setting can still be described by a line element,
presupposing a Riemannian structure of space-time, is therefore
unjustified. Detailed discussions of the underlying modifications of
contributions to the Hamiltonian constraint from background and perturbations
show that the implicit assumption of unmodified brackets, and thus Riemannian
structures, is inconsistent \cite{NonCovDressed}.

For a consistent space-time structure, the gauge behavior of the classical
theory must remain intact, even while it may be modified and subject to
quantum effects. In general, this condition requires anomaly freedom, such
that the same number of physical degrees of freedom as in classical gravity
are realized in a modified version. If this condition is violated, the
modified theory cannot have the correct classical limit owing to a
discontinuity in the number of degrees of freedom. An anomalous modification
or quantization of gravity does not permit a semiclassical or effective
treatment by line elements in any form because it is incompatible with the
gauge structure of space-time.

A formal statement of the condition that the gauge behavior remain intact is
the existence of closed Poisson brackets of $H[N]$ and $D[M^i]$ for all relevant $N$
and $M^i$, depending on whether one considers the full theory or a restricted
version such as a midisuperspace model. This condition allows for possible
quantum corrections in the structure functions of the gauge algebra, given in
the case of gravity by the inverse spatial metric $q^{ij}$ as it appears in
\begin{equation}
 \{H[N_1],H[N_2]\}= D[\beta(q,p) q^{ij}(N_1\nabla_jN_2-N_2\nabla_jN_1)]
\end{equation}
with a possible modification function $\beta(q,p)$ on phase space.
We have the
classical space-time structure if $\beta=\pm1$, giving two possible choices of
the signature of a classical 4-dimensional metric, where $\beta=1$ for
Lorentzian-signature space-time and $\beta=-1$ for 4-dimensional
Euclidean-signature space. (In each case, the name refers only to the
signature and does not imply flatness.)

We have a consistent non-classical space-time structure if the brackets are
closed such that $\beta\not=\pm 1$. The modification function $\beta$
determines the structure functions of hypersurface-deformation brackets in the
modified theory. Modified structure functions, in turn, show via
(\ref{deltaN}) how lapse and shift transform and whether it is possible to
find suitable field redefinitions of these fields that can be used in a proper
effective line element as discussed in detail in \cite{EffLine}.

As we have seen in the present section, suitable
transformations of lapse and shift as components of the space-time metric
require knowledge of the structure functions of $H[N]$ and $D[M^i]$. If the
brackets do not close, as in the dressed-metric approach, there are no
meaningful transformations of lapse and shift and it is impossible to
construct a valid structure of space-time. Such a structure exists only in
anomaly-free modifications of the constraints. However, the condition of
anomaly-freedom is not sufficient if it does not imply a clear modification of
the structure function of hypersurface-deformation brackets, for instance in
cases in which the constrained system is reformulated before it is modified or
quantized. An example for such an approach is given by a partial
Abelianization of the constraints \cite{LoopSchwarz}, to which we turn next.

\section{Spherical symmetry}
\label{s:SphSymm}

An instructive set of examples is given by spherically symmetric space-time
geometries with line element
\begin{equation}
 {\rm d}s^2=-N^2{\rm d}t^2+L^2({\rm d}x+M{\rm d}t)^2+S^2({\rm
  d}\vartheta^2+\sin^2\vartheta{\rm d}\varphi^2)
\end{equation}
where $N$, $L$, $M$ and $S$ are functions of $t$ and $x$. Together with the
momenta $p_L$ and $p_S$ of $L$ and $S$, respectively, the components $L$ and
$S$ of the spatial metric in classical general relativity are subject to the
Hamiltonian constraint
\begin{equation}
 H[N]= \int_{-\infty}^{\infty} N\left(-\frac{p_Lp_S}{S}+
   \frac{Lp_L^2}{2S^2}+ 
\frac{(S')^2}{2L} +\frac{SS''}{L}-
 \frac{SS'L'}{L^2} - \frac{L}{4} \right){\rm d}x
\end{equation}
and the diffeomorphism constraint
\begin{equation}
 D[\epsilon]= \int_{-\infty}^{\infty} \epsilon \left(p_S
   S'-Lp_L'\right){\rm d}x\,.
\end{equation}
The relevant bracket with a stucture function is given by
\begin{equation} \label{HH}
\{H[N_1],H[N_2]\}= D[L^{-2} (N_1N_2'-N_2N_1')\,.
\end{equation}

\subsection{Reformulating the constrained system}

In \cite{LoopSchwarz}, a reformulation of the constraints has been suggested
that can remove the structure function and even partially Abelianize the
brackets. Instead of $H[N]$, this reformulation uses the linear combination
\begin{equation} \label{HD}
H[2P S'/L]+D[2Pp_L/(SL)]= \int_{-\infty}^{\infty} P
 \frac{{\rm d}}{{\rm d}x} \left(-\frac{p_L^2}{S}+ \frac{S(S')^2}{L^2}-S
   \right){\rm d}x
\end{equation}
of Hamiltonian and diffeomorphism constraints. Specifically, the combination
replaces $H[N]$ with a new constraint whose integrand (except for the
multiplier $P$) is a complete derivative. Imposing (\ref{HD}) as a constraint
therefore requires that the parenthesis in this expression equals a constant,
$C_0$. The same condition can be expressed by the alternative constraint
\begin{equation}
 C[Q]= \int_{-\infty}^{\infty} Q \left( -\frac{p_L^2}{S}+
   \frac{S(S')^2}{L^2}-S- C_0\right){\rm d}x\,.
\end{equation}
(The constant can be related to boundary values.) Because $C[Q]$ depends
neither on $p_S$ nor on spatial derivatives of $L$, it is easy to see that two
such constraints always have a vanishing Poisson bracket, unlike two
Hamiltonian constraints.  Together with the original diffeomorphism
constraint, we have brackets
\begin{equation} \label{CD}
 \{C[Q],D[\epsilon]\}=-C[(\epsilon Q)'] \quad,\quad \{C[Q_1],C[Q_2]\}=0
\end{equation}
free of structure functions. Therefore, it may be expected that using the
reformulated constraints greatly simplifies the quantization procedure or the
derivation of viable modifications.

However, the reformulation has made use of metric-dependent coefficients
$S'/L$ and $p_L/(SL)$ in (\ref{HD}). In general, it is not clear whether these
coefficients will be subject to quantum corrections, in which case it may be
difficult or impossible to reconstruct valid hypersurface-deformation brackets
with the correct classical limit from a quantization or modification of the
system (\ref{CD}). The non-trivial nature of this question has been shown in
\cite{SphSymmCov} and the related \cite{GowdyCov}, where examples were
presented in which (\ref{CD}) can easily be modified while no
hypersurface-deformation brackets can be reconstructed at all or only in
modified form.

For instance, the modification
\begin{equation} \label{Cf}
 C_f[Q]= \int_{-\infty}^{\infty} Q \left( -\frac{f(p_L)^2}{S}+
   \frac{S(S')^2}{L^2}-S- C_0\right){\rm d}x
\end{equation}
with a free function $f(p_L)$, such as $\sin(\ell p_L)/\ell$ where $\ell$ is a
suitable length scale analogous to (\ref{HolFriedmann}), and an unchanged
$D[\epsilon]$ maintains the brackets (\ref{CD}) and is therefore anomaly-free
in the reformulated system. By reverting the steps undertaken in (\ref{HD}),
it can be seen that (\ref{Cf}) corresponds to the modified Hamiltonian
constraint
\begin{equation} \label{Hf}
 H_f[N]= \int_{-\infty}^{\infty}
 N\left(-\frac{p_S}{S}\frac{{\rm 
       d}f(p_L)}{{\rm d}p_L}+ 
   \frac{Lf(p_L)}{2S^2}+ 
\frac{(S')^2}{2L} +\frac{SS''}{L}-
 \frac{SS'L'}{L^2} - \frac{L}{4} \right){\rm d}x \,.
\end{equation}
This modification of the Hamiltonian constraint, which has already been found
in \cite{JR}, also turns out to be anomaly-free, but with a modified bracket
\begin{equation}
 \{H_f[N_1],H_f[N_2]\} = D[\beta(p_L)L^{-2} (N_1N_2'-N_2N_1')]
\end{equation}
where
\begin{equation}
 \beta(p_L) = \frac{1}{2} \frac{{\rm d}^2f}{{\rm d}p_L^2}\,.
\end{equation}
The modified structure function is an example of signature change because
$\beta$ is negative around any local maximum of $f$.

If spherically symmetric gravity is coupled to a scalar field, the partial
Abelianization of \cite{LoopSchwarz} is still available and can be modified as
in (\ref{Cf}). However, in this case there is no consistent set of
hypersurface-deformation generators \cite{SphSymmCov}. Therefore, the modified
theory is formally consistent but not geometrical: Its solutions cannot be
described by Riemannian geometry or effective line elements, even after a
field redefinition. This problem poses a significant challenge to loop
quantization because an application only to vacuum models would be too
restrictive. Moreover, the problem is broader because polarized Gowdy models,
which can also be partially Abelianized, do not admit a consistent set of modified
hypersurface-deformation brackets \cite{GowdyCov}. So far, therefore,
midisuperspace models with local physical degrees of freedom cannot be
described geometrically in the presence of loop modifications.

\subsection{Non-bijective canonical transformation}

To circumvent this problem, \cite{CovPol} proposed a modification of
spherically symmetric gravity based on a non-bijective canonical transformation:
\begin{equation}
 p_L= \frac{\sin(\ell \tilde{p}_L)}{\ell} \quad,\quad
 L= \frac{\tilde{L}}{\cos(\ell \tilde{p}_L)}\,.
\end{equation}
The transformation can be applied to the Abelianized constraint $C[Q]$ or to
the Hamiltonian constraint by inserting $p_L(\tilde{p}_L)$ and
$L(\tilde{L},\tilde{p}_L)$ in their classical expressions. (The diffeomorphism
constraint is not modified by this transformation.) Terms depending on $p_L$
in $C[Q]$ are then modified as before in (\ref{Cf}) with a specific version of
$f(p_L)$, and there are new modifications in the $L$-term. As postulated in
\cite{CovPol}, this procedure, based on a canonical transformation, might be
able to preserve covariance of the classical theory even in the presence of a
scalar field, and yet allow room for new quantum effects because of the
non-bijective nature of the canonical transformation.

Unfortunately, this hope remains unfulfilled precisely because the transformation
is not bijective \cite{NonCovPol}. In particular, the bijective nature breaks
down at hypersurfaces defined by $\ell p_L=\pm 1$ or
$\ell\tilde{p}_L=(n+1/2)\pi$, and $p_L$ as well as $\tilde{p}_L$ are spatial scalars
but not space-time scalars. Therefore, while the transformation preserves
symmetries of the classical theory when it can be restricted to regions of
phase space in which it is bijective, these regions themselves are defined in
terms that are not space-time covariant. The resulting theory is not covariant.

For the same reason, $\tilde{p}_L$ not being a space-time scalar, the variable
$\tilde{L}$ introduced by the canonical transformation does not have the same
behavior as $L=\tilde{L}/\cos(\ell\tilde{p}_L)$ under space-time
transformations. As a consequence, $\tilde{L}$ cannot be used in
a space-time line element based on $\tilde{L}^2{\rm d}x^2$. A meaningful
effective line element is obtained only after a suitable field redefinition
that leads to a function of $\tilde{L}$ with the correct transformation
properties. Since we already know that $\tilde{L}$ has been derived from such
a function, $L$, the field redefinition simply sends us back from $\tilde{L}$
to $L$ in regions in which the canonical transformation is invertible, undoing
the modification of the theory in such regions. (More systematically, such a
field redefinition can be derived using methods introduced in \cite{Absorb}.)
In these regions, exact classical solutions without any modifications are produced,
but different regions are connected along hypersurfaces (again, given by
$\ell p_L=\pm 1$ or $\ell\tilde{p}_L=(n+1/2)\pi$) that are not covariantly
defined. Since these hypersurfaces refer to fixed values of certain components
of extrinsic curvature, their positions in space-time depend on choices of
coordinates and spatial slicings.

In particular, slicings with large $p_L\sim 1/\ell$ exist even in flat
space-time, and therefore violations of covariance in this model cannot be
considered a ``large-curvature effect.''  These violations can occur at low
space-time curvature (in an invariant meaning), and therefore the model cannot
be considered a permissible model of quantum gravity that would have
non-standard geometrical features only at the Planck scale. The model could be
permissible only if it were combined with a mechanism that somehow prevents
one from choosing slicings that lead to large extrinsic curvature $p_L$. But
preventing such slicings (or any slicing) from being allowed requires
violations of covariance that are hard to reconcile with the application of line
elements, even if they were used only in low-curvature regions.

\subsection{Bijective canonical transformation}

As discussed in more detail in \cite{NonCovPol}, the application of canonical
transformations makes an analysis of space-time structure rather non-trivial
even if the transformation is bijective. A bijective canonical transformation
from $(L,p_L)$ to some $(\tilde{L},\tilde{p}_L)$ may well be such that all
possible values of $p_L$ are mapped to a finite range of $\tilde{p}_L$. One
could then conclude that the transformed theory resolves singularities if
$\tilde{p}_L$, interpreted as some curvature expression in the new theory,
remains bounded. However, the new theory has been obtained by applying a
bijective canonical transformation that cannot modify the physics of classical
spherically symmetric models.

The answer to this conundrum relies on effective line elements. For a
transformation with a significantly modified $\tilde{p}_L$ to be canonical,
$\tilde{L}$ must also be modified compared with $L$. Then, the structure
function in (\ref{HH}) is modified when expressed in terms of $\tilde{L}$
instead of $L$, and solutions of the transformed theory cannot be interpreted
directly in terms of a line element where $\tilde{L}$ directly takes the place of
$L$. An effective line element, derived again as in \cite{Absorb}), requires
an undoing of the canonical transformation for a valid coefficient of ${\rm
  d}x^2$, sending us back to the classical theory in its geometrical
interpretation.

Models of loop quantum gravity are not obtained by bijective canonical
transformations and could lead to new physics. However, the example of a
bijective canonical transformation demonstrates that predictions can be
reliable only if a proper effective line element is derived. Unfortunately,
this task is rarely performed in phenomenological studies of models of loop
quantum gravity. In several proposals, as in the dressed-metric approach, it
is even impossible to construct an effective line element because they do not
amount to consistent modifications of the crucial bracket (\ref{HH}) that
determines the structure of space-time.

\section{Homogeneity in Schwarzschild space-time}
\label{s:Hom}

It is well known that a spatially homogeneous geometry of Kantowski--Sachs
type \cite{KS}, with line element
\begin{equation} \label{KS}
 {\rm d}s^2= -N(t)^2{\rm d}t^2+ a(t)^2{\rm d}x^2+ b(t)^2
 \left({\rm 
     d}\vartheta^2+ \sin^2\vartheta{\rm d}\varphi^2\right)
\end{equation}
is realized in the Schwarzschild interior:
In the (almost) standard version
\begin{equation}
 {\rm d}s^2 =- (1-2M/r) {\rm d}\tilde{t}^2+\frac{{\rm d}r^2}{2M/r-1}+ r^2 \left({\rm 
     d}\vartheta^2+ \sin^2\vartheta{\rm d}\varphi^2\right)
\end{equation}
of the Schwarzschild line element, $\tilde{t}$ is a time coordinate only for
$r>2M$, outside of the horizon. For $r<2M$, the coordinate $r$ may be used as
time while $\tilde{t}$ contributes to a positive, spacelike part of the line
element. Indicating the modified roles of the coordinates in the notation, we
define $t=r$ and $x=\tilde{t}$ for $r<2M$, such that the line element turns
into
\begin{equation}
 {\rm d}s^2 =- \frac{{\rm d}t^2}{2M/t-1}+ (2M/t-1) {\rm d}x^2+ t^2 \left({\rm 
     d}\vartheta^2+ \sin^2\vartheta{\rm d}\varphi^2\right)
\end{equation}
for $t<2M$. A suitable identification of $N(t)$, $a(t)$ abd $b(t)$ shows that
this line element is of the general form (\ref{KS}).

The coordinates $t$ and $x$ determine a homogeneous spacelike slicing in the
interior of Schwarzschild space-time.
It is therefore possible to apply minisuperspace quantizations to
the interior region.  However, such models do not show how a modified quantum
interior may be connected to an inhomogeneous exterior, and they do not reveal
properties of space-time structure (let alone physical processes such as
occasionally hypothesized explosions of black holes).

\subsection{Timelike homogeneity of exterior static solutions}

A complex canonical transformation $A=ia$ and $p_A=-ip_a$ together with $n=iN$
in (\ref{KS}) implies a Kantowski--Sachs line element of the form
\begin{equation} \label{KSComplex}
 {\rm d}s^2= n(t)^2{\rm d}t^2- A(t)^2{\rm
   d}x^2+ b(t)^2 
 \left({\rm  
     d}\vartheta^2+ \sin^2\vartheta{\rm d}\varphi^2\right)\,.
\end{equation}
The complex transformation has the same effect as crossing the horizon in the
Schwarzschild geometry: It flips the roles of $t$ and $x$ as time and space
coordinates. Defining $X=t$ and $T=x$, the transformed line element
(\ref{KSComplex}) takes the form
\begin{equation} \label{KSComplex}
 {\rm d}s^2= -A(X)^2{\rm d}T^2+n(X)^2{\rm d}X^2+ b(X)^2 
 \left({\rm  
     d}\vartheta^2+ \sin^2\vartheta{\rm d}\varphi^2\right)\,.
\end{equation}
The exterior Schwarzschild line element
\begin{equation}
 {\rm d}s^2 = -(1-2M/X) {\rm d}T^2+ \frac{{\rm d}X^2}{1-2M/X}+ X^2 \left({\rm 
     d}\vartheta^2+ \sin^2\vartheta{\rm d}\varphi^2\right)
\end{equation}
with $X>2M$ is now of this general form. In particular, the coordinates $T$
and $X$ determine a homogeneous timelike slicing in the exterior.
Methods of minisuperspace quantization can therefore be applied even to
inhomogeneous geometries \cite{Transfig}, possibly leading to modified
space-time structures.

\begin{figure}
\begin{center}
\includegraphics[width=12cm]{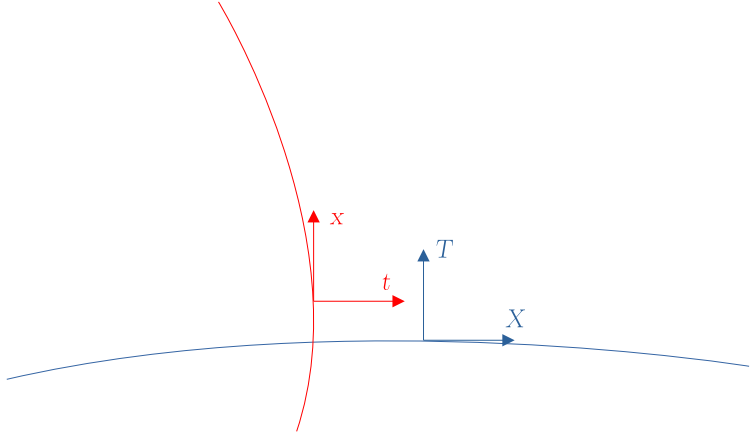}
\caption{A homogeneous timelike slicing with coordinates $(t,x)$ and an
  inhomogeneous spacelike slicing with coordinates $(T,X)$, both in the same static
  spherically symmetric space-time.
  \label{f:Slicings}}
\end{center}
\end{figure}

Symmetries of individual space-time solutions such as homogeneity, as opposed
to general covariance which relates different solutions of the underlying
partial differential equations, are built into the setup of the
model. Therefore, they are preserved by minisuperspace quantization. Timelike
homogeneity then remains intact for any modified dynamics in this setting. As
shown in Fig.~\ref{f:Slicings}, timelike homogeneity with the given number of
degrees of freedom, in turn, implies the existence of a static spherically
symmetric configuration if the resulting theory is covariant and slicing
independent (described by a meaningful line element). Since the black-hole
analysis of \cite{Transfig} is based on line elements and refers to notions of
Riemannian geometry, such as horizons, curvature scalars or Penrose diagrams,
slicing independence is one of the ingredients of the construction and need
not be assumed independently. It must therefore be possible to formulate the
same physics claimed in \cite{Transfig} for a homogeneous timelike slicing
also within a covariant spherically symmetric theory, restricted to static
solutions.

Covariant versions of spherically symmetric gravity models and their static
solutions are under good control, thanks to work on dilaton gravity
\cite{Strobl,DilatonRev} and its generalizations
\cite{NewDilaton,DilatonHorndeski}. It is therefore possible to check whether
a proposed modification of the homogeneous timelike slicing has a chance of
corresponding to a covariant theory.

\subsection{Line elements}

Timelike homogeneity with modified dynamics leads to a formal line element
\begin{equation} \label{stilde}
 {\rm d}s^2= \tilde{n}(t)^2{\rm d}t^2- \tilde{A}(t)^2{\rm 
   d}x^2+ \tilde{b}(t)^2 
 \left({\rm  
     d}\vartheta^2+ \sin^2\vartheta{\rm d}\varphi^2\right)
\end{equation}
if solutions $\tilde{n}$, $\tilde{A}$ and $\tilde{b}$ are simply inserted in
the classical line element. Since properties of space-time transformations
have not been checked at this point, there is no guarantee that (\ref{stilde})
presents a proper effective line element.

Assuming that the Kantowski--Sachs-like (\ref{stilde}) is a proper line
element that describes a slicing-independent theory, it is equivalent to the
Schwarzschild-like 
\begin{equation} \label{sKL}
{\rm d}s^2= -K(X)^2{\rm d}T^2+ L(X)^2 {\rm d}X^2+ S(X)^2
 \left({\rm   d}\vartheta^2+\sin^2\vartheta{\rm d}\varphi^2\right)
\end{equation}
where $X=t$, $T=x$ and 
\begin{equation} \label{AK}
 \tilde{A}=K\quad,\quad \tilde{b}=S\quad,\quad \tilde{n}=L\,.
\end{equation}
By construction, the coefficients in (\ref{sKL}) depend on $X$ but not on
$T$. The line element therefore presents a static solution in a spherically
symmetric model, subject to some modified dynamics because $K$, $L$ and $S$
are only Schwarzschild-like but not exactly of Schwarzschild form if the
dynamics of the underlying homogeneous model is modified. 

If the assumption of covariance, made implicitly in \cite{Transfig}, is
justified, (\ref{sKL}) must be a solution of a $1+1$-dimensional gravity model
in terms of time and space coordinates $(T,X)$. Such theories are under strong
control: All local covariant theories of this midisuperspace form are known as
generalized dilaton gravity models \cite{NewDilaton}. (Their equivalence to
Horndeski theories in $1+1$ dimensions has been shown in
\cite{DilatonHorndeski}.) While several free functions exist in this general
setting to specify the dynamics, for instance through an action, they can only
depend on the variable analogous to our field $S$. Loop quantum cosmology
applied to the homogeneous timelike slicing, however, implies modifications
that do not fulfill this condition: Such minisuperspace modifications depend
non-linearly on momenta $p_{\tilde{A}}$ and $p_{\tilde{b}}$, which are linear
combinations of ${\rm d}\tilde{A}/{\rm d}t$ and ${\rm d}\tilde{b}/{\rm d}t$
that, according to (\ref{AK}), are translated to $\partial K/\partial X$ and
$\partial S/\partial X$ in the spherically symmetric slicing. Therefore, no
holonomy modified dynamics of Kantowski--Sachs-style models can be part of a
covariant space-time theory \cite{Disfig}.

\section{Conclusions}

We have discussed the main constructions that were supposed to circumvent
difficulties in earlier applications of loop quantization to inhomogeneous
models. Instead of solving older problems, however, these constructions led to
no-go results for covariance in models of loop quantum gravity.
A complete understanding of covariance in any given model is important not
only to demonstrate its consistency, but also to evaluate possible
observational implications of the underlying theory. For instance, if one
neglects to identify suitable space-time structures for a model of modified or
quantum gravity, one could be led to posing initial conditions at an
inadmissible place where there is, in fact, no meaningful version of time. A
detailed space-time analysis may well show other regions in which initial values
could reliably be posed, but the altered location, perhaps at a different
range of curvature values, would affect implied phenomenological
effects. Addressing such questions requires an understanding of different ways
in which covariance can be violated, which we compare in the next
subsection. The final two subsections will discuss general implications for
loop quantum gravity and a brief outlook on covariance in other approaches.

\subsection{Comparison of different violations of covariance}

The examples reviewed in the preceding sections show different ways in which
covariance can be violated in models of loop quantum gravity. The
dressed-metric approach, just as hybrid loop quantum cosmology, is based on the
incorrect assumption that background and perturbations can be quantized or
modified independently in an inhomogeneous model. This assumption ignores a
crucial feature of space-time and covariance, according to which background
and perturbative transformations form a semidirect product but not a direct
one as an independent treatment would require. The fundamental nature of this
property implies that covariance is completely broken in these models, which
are therefore inconsistent as description of (quantum) space-time.

As usual, one may expect that space-time is non-classical at large curvature
and may exhibit properties different from classical space-time. However, this
expectation does not redeem quantum models that violate covariance unless they
can demonstrate that the classical properties are recovered in a suitable
classical limit. Moreover, the dressed-metric and the hybrid approach both
refer to features of classical space-time, such as line elements or curvature
perturbations, even close to the Planck curvature.

The inconsistency of these approaches is rooted not so much in possible
modifications of classical space-time properties near the Planck curvature,
but rather in the unquestioned (and often implicit) application of classical
space-time ingredients for an analysis in this regime. For a model to be
consistent, such an assumption must be justified, but this crucial step has
not been attempted in the dressed-metric and hybrid approaches. There is
therefore reason to doubt the validity of these constructions and their
implications.

The technical observation that a key property of classical space-time is
violated, given by the semidirect-product nature of transformation, serves as
a concrete property that turns this doubt into a proof that the models are
inconsistent, not only in the Planck regime but to any order in a
semiclassical expansion by $\hbar$ or $\ell_{\rm P}$. Consistency is recovered
only in the strict limit of $\hbar\to0$, just because we happen to know that
the classical theory is covariant and has solutions that can be described by
line elements. In such modifications, there is a strong discontinuity at $\hbar=0$ in
geometrical structures, seen as an $\hbar$-dependent family of
modifications. In practice, this discontinuity translates into low-curvature
physical problems, as discovered in the case of black-hole models of loop
quantum gravity in \cite{TransCommAs,LoopISCO}.

Similarly, the original attempt in \cite{Transfig} to describe the
inhomogeneous Schwarzschild exterior by homogeneous models, using timelike
slicings in a static geometry, was based on an untested assumption that is
true in classical space-time but may be violated in the presence of quantum
modifications. The description of inhomogeneity in this case is different
from the preceding example because it is non-perturbative in a spacelike
slicing. Here, homogeneous and inhomogeneous configurations do not appear as
background and perturbations, but rather as models of a single space-time
geometry using two different slicings.

Classically, any slicing gives an equivalent description of the full geometry,
but this need not be the case once equations have been modified, in contrast to
what has implicitly been assumed in \cite{Transfig}. The good control on
covariant local theories for spherically symmetric dynamics makes it possible
to test and invalidate this assumption. Again, it is the application of line
elements in \cite{Transfig} even in the presence of quantum modifications that
makes it possible to demonstrate inconsistency. It is not necessary to assume
additional classical features in the inconsistency proof, beyond properties
that have already been used in \cite{Transfig}, explicitly or implicitly.

Models that work directly with spherically symmetric inhomogeneity usually
tread more carefully because the appearance of first-class constraints is
explicit. A consistent quantization or modification then requires that the
first-class nature be preserved, that is, that there be no anomalies, in order
to avoid spurious degrees of freedom or over-constraining the theory. However,
even in an anomaly-free modification, the structure of space-time and geometry
may remain unclear without further analysis. Here, our remaining two examples
are relevant, given by different modifications implemented for reformulated,
partially Abelianized constraints and modification through a non-bijective
canonical transformation, respectively. These modifications are anomaly-free
and therefore consistent in a formal sense used for general constrained
systems. Nevertheless, they turn out to violate covariance in different ways,
even though the papers in which they have been proposed go on and analyze
their solutions by standard line elements.

\subsection{Covariance crisis of loop quantum gravity}

As we just saw, a crucial ingredient of proofs of inconsistency and
non-covariance in models of loop quantum gravity focuses on the application of
line elements used routinely to evaluate solutions of modified equations
in canonical gravity. Since modifications of canonical equations need not
preserve covariance, even if they may remain formally consistent and
anomaly-free, line elements are rendered meaningless. It might therefore be
possible to evade some of the no-go results by foregoing line elements or
related and more advanced methods, such as Penrose diagrams. In principle, a
physical analysis would still be possible, at least in the anomaly-free case,
by expressing solutions of anomaly-free modified equations in terms of
suitable canonical observables.

However, this option is rarely exercised in interesting models because of the
complicated nature of deriving strict observables, compared with the simple
procedure of modifying coefficients in a formal line element. And if such
an analysis could be performed, it would not be clear in which sense solutions
of the modified theory could still be considered geometrical, even when
quantum modifications are very small, or more practically, how one would
define the horizon of a black hole or curvature perturbations for cosmology in
the absence of geometry. The important covariant form of general
relativity and its geometrical nature would be a mere accident of the
classical theory, rather than a fundamental property of gravity that could be
extended even to the tiniest of corrections. While requiring a geometrical
nature for quantum gravity may be largely a matter of taste, it also has
practical implications because most of the gravitational methods and
definitions that we know and understand are based on geometry.

A few additional ways might remain to solve these deep problems.  First, in the
context of Section~\ref{s:Hom}, non-local effects might help because they
would evade the strong control on possible covariant theories with spherical
symmetry. However, the underlying analysis of minisuperspace dynamics in
\cite{Transfig} implicitly assumes locality because there is a single momentum
for each classical metric or triad component. If one were to try non-locality
in order to solve the covariance problem in models of loop quantum gravity,
the entire formalism used up until now would have to change, even in
minisuperspace models.  Moreover, non-locality is often pathological and there
is no indication that loop quantization could lead to more controlled
situations.

Secondly, one may try to understand non-Riemannian space-time structures as
they would be implied by modified hypersurface-deformation brackets
($\beta\not=\pm 1$). In some (but not all) cases, these modified geometries
can be described by an effective Riemannian line element after suitable field
redefinitions. At present, such models, analyzed recently in
\cite{EffLine,DefSchwarzschild,DefSchwarzschild2,DefGenBH}, are the only
well-defined descriptions of geometries that may incorporate quantum
modifications.  If suitable field redefinitions exist, strict effective line
elements are available, but in the presence of holonomy modifications they
generically imply signature change at high curvature.

There has been progress on constructing anomaly-free versions of the
Hamiltonian constraint directly at the operator level in various versions of
loop quantum gravity
\cite{TwoPlusOneDef,TwoPlusOneDef2,AnoFreeWeak,AnoFreeWeakDiff,OffShell,ConstraintsG}. These
constructions do not directly refer to symmetry-reduced models but, for now,
implement restrictions of general ingredients such as the spatial dimension,
the local gauge group, or the signature of gravity. In this approach, progress
is usually made by reformulating the constraints, simplifying their brackets
in a way that is conceptually similar to partial Abelianizations discussed in
Section~\ref{s:SphSymm}. As in this case, the successful construction of
anomaly-free reformulated constraints does not immediately reveal whether they
describe a consistent structure of space-time or covariance.

\subsection{Lessons for other approaches}

Background independence implies that space-time structure must be derived in
some way and cannot be presupposed.  We should not simply assume that
inserting modified solutions in classical-type line elements is consistent.
As a consequence, quantum gravity may not be ``geometrical'' as we understand
it from general relativity. In the main body of this paper we discussed how
the canonical nature of loop quantum gravity gives access to powerful
space-time methods, based on algebra, that can be used to rule out many models
that might otherwise look reasonable.

It is not easy to see whether there may be possible analogs of our results in
alternative approaches to quantum gravity if they are not
canonical. Nevertheless, we are able to draw several lessons of general form.
First, non-canonical theories do not directly aim to quantize generators of
hypersurface deformations, but it should still be of interest to construct
them and consider their properties in order to facilitate a space-time
analysis. Instead of using these generators, covariance is often expressed in
terms of coordinate choices or embeddings of discrete structures, but these
ingredients do not directly refer to the actual degrees of freedom of
gravity. Moreover, the explicit application of these space-time ingredients
reduces the freedom in formulating suitable modifications of space-time
structures if they are called for by modified dynamics.

Secondly, the no-go results we have encountered are very general. In
particular, they do not require a specific form of modifications but only
qualitative features related to discreteness, such as bounded modification
functions with local maxima. They should therefore be expected to be largely
independent of the specific approach.  Even though they have been derived for
canonical quantum gravity, the no-go results can be applied to any modified
cosmological dynamics that can be presented in canonical form, even if it has
been derived from a non-canonical approach. It would be interesting to see how
other approaches might be able to circumvent our no-go theorems, for instance
by requiring new quantum degrees of freedom or specific non-local
behaviors. (For an example of non-local effects derived for effective actions,
see \cite{EffAcRecons}.)

Finally, hypersurface-deformation generators make it possible to analyze
different space-time structures because they express geometrical properties
through algebra. It is easier to control possible modifications or
deformations of algebras (or algebroids), compared with geometrical
structures. The strong algebraic background of canonical gravity is therefore
the main reason why it is possible to analyze space-time structures in detail
with canonical methods. Non-canonical approaches are often viewed as
preferable because they can provide a direct 4-dimensional space-time picture,
at least heuristically. However, this proximity to the standard 4-dimensional
formulation of classical gravity also implies that hidden assumptions about
the underlying geometry may easily and unwittingly be incorporated in a
specific approach. As shown in the present paper, even canonical approaches
are not immune to such hidden assumptions, but they also provide strong
methods to spot and test unjustified assumptions.

\section*{Acknowledgements}

This paper is based on a talk given to the quantum gravity group at Radboud
University, Nijmegen. The author is grateful to Renate Loll for an invitation
and insightful questions, as well as to Jan Ambj\o rn, Suddhasattwa Brahma and
Timothy Budd for discussions. This work was supported in part by NSF grant
PHY-1912168.


\end{document}